\documentclass[sn-basic]{sn-jnl}

\usepackage{graphicx,subcaption,caption}%
\usepackage{multirow,tikz}%
\usepackage{amsmath,amssymb,amsfonts}%
\usepackage{amsthm}%
\usepackage{mathrsfs}%
\usepackage[title]{appendix}%
\usepackage{xcolor}%
\usepackage{textcomp}%
\usepackage{manyfoot}%
\usepackage{booktabs}%
\usepackage{algorithm}%
\usepackage{algorithmicx}%
\usepackage{algpseudocode}%
\usepackage{listings}%

\begin{document}

\title{
\begin{center}
    Assimilating rough features:
\end{center} 
\vspace{-3mm}
A data-driven framework to infer rough wall properties from sparse experimental data}

\author*[1]{\fnm{Martina} \sur{Formichetti}}\email{martina.formichetti@soton.ac.uk}

\author[1]{\fnm{Uttam} \sur{Cadambi Padmanaban}}

\author[2]{\fnm{Ping} \sur{He}}

\author[1]{\fnm{Sean}\sur{Symon}}
\author[1]{\fnm{Bharathram}\sur{Ganapathisubramani}}

\affil[1]{\orgdiv{Department of Aero and Astro Engineering}, \orgname{University of Southampton}, \orgaddress{\city{Southampton}, \country{UK}}}

\affil[2]{\orgdiv{Department of Aerospace Engineering}, \orgname{Iowa State University}, \orgaddress{\city{Ames}, \country{USA}}}


\abstract{Surface roughness influences turbulent boundary layers (TBLs) primarily through the roughness function $\Delta U^+$ and the equivalent sand-grain roughness height \(k_s\). Direct determination of \(k_s\) typically requires detailed velocity and wall-shear stress measurements, which are often impractical. As an alternative, this study presents a data assimilation framework that modifies a smooth-wall Reynolds-Averaged Navier–Stokes (RANS) baseline to match sparse rough-wall particle image velocimetry (PIV) data in the fully rough regime. Through this approach, secondary variables such as the friction velocity, \(u_\tau\), and \(k_s\) can be inferred from the assimilated flow fields. The assimilated TBL reproduces experimental velocity profiles within 1\% and predicts friction velocity within 1-6\% of the experimental measurements. Furthermore, the \(k_s\) values inferred from the assimilation also match the experimental data up to 1\%. These results demonstrate the potential of data assimilation as a cost-effective alternative to high-fidelity methods and support the generalisation of the framework to model streamwise-varying roughness by treating \(k_s\) as a function of fetch length.}

\keywords{Data assimilation, Turbulent boundary layers, Turbulence, Roughness}



\maketitle

\section{Introduction}\label{sec1}
\begin{figure}[ht]
    \centering
    \includegraphics{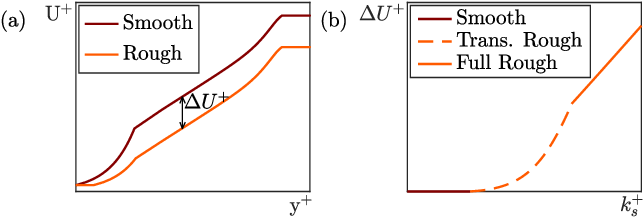}
    \caption{Definition of roughness function, $\Delta U^+$, as  (a) the difference between smooth and rough mean velocity log-profiles in viscous units, and (b) as a function of roughness Reynolds number, $k_s^+$}
    \label{fig1}
\end{figure}

TBLs over rough walls play a central role in a wide range of engineering and geophysical flows, from aerodynamic drag on aircraft and ship hulls \citep{Whitmore2002,Schultz2007} to the dynamics of atmospheric boundary layers \citep{BouZeid2020}. Surface roughness influences these flows primarily through the roughness function, $\Delta U^+$, and the equivalent sand-grain roughness height, $k_s$. As shown in Fig.~\ref{fig1}a, the presence of roughness results in a downward shift of the mean velocity profile in viscous units (U$^+$ vs y$^+$) compared to a smooth-wall baseline. This shift, defined as the roughness function $\Delta U^+$, effectively represents a momentum deficit resulting from the presence of roughness in the flow. The magnitude of $\Delta U^+$ is a function of the roughness Reynolds number, $k_s^+ = k_s u_\tau / \nu$, where $u_\tau$ is the friction velocity, and $\nu$ is the kinematic viscosity of the fluid. There are three distinct roughness regimes as illustrated in Fig.~\ref{fig1}b: aerodynamically smooth, transitionally rough, and fully rough \citep{Jimenez2004}. In the logarithmic region, this relationship is expressed as

\begin{equation}
\Delta U^+ = \frac{1}{\kappa} \ln(k_s^+) + B_s - B_R = \frac{1}{\kappa} \ln(C_s k_s^+),
\label{eq1}
\end{equation}
where $B_s$ is the smooth-wall intercept of the log law, typically $4\leq B_s\leq5$, and $B_R$ is the rough-wall intercept, which for fully rough flow is $B_R \approx8.5$ \citep{Schlichting2016}. The constant $C_s$ represents a grouping of these constants defined as
\begin{equation} \label{eq:Cs}
    C_s = \exp(\kappa(B_s - B_R)).
\end{equation}
Equations~\ref{eq1} and \ref{eq:Cs} are also used to represent the transitionally rough regime by tuning the constant $C_s$. This allows the model to account for the varying transitional behaviours of different surfaces, ranging from the inflectional shift observed by \cite{Nikuradse1933} for uniform sand to the monotonic transition described by \cite{Colebrook1937} for commercial pipes.

\begin{figure}[ht]
    \centering
    \includegraphics[width = 0.8\linewidth]{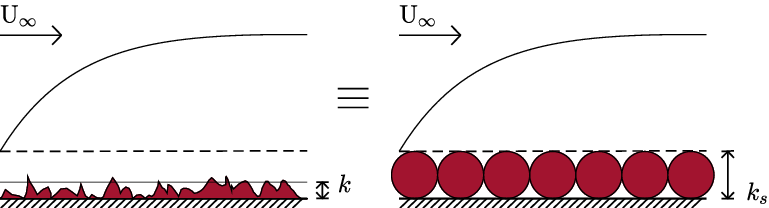}
    \caption{Equivalent sand-grain roughness height definition. Figure adapted from \citet{Formichetti2025}}
    \label{fig2}
\end{figure}

It is important to emphasise that $k_s$ is not a physical length scale of the geometry. Rather, as indicated in Fig.~\ref{fig2}, it is a length scale used to characterise rough flows that is equivalent to the physical height of an idealised sand-grain type of roughness that would affect the flow in an equivalent way to the arbitrary roughness. This formulation allows complex surfaces to be described in terms of an idealised equivalent, making $k_s$ a fundamental quantity for both experimental and computational studies. An example of the usage of $k_s$ in practical applications can be found in \citet{Schultz2009,Flack2014}.

Traditionally, determining \(k_s\) requires extensive experimental campaigns to measure velocity profiles and wall shear stress, from which \(\Delta U^+\) is computed. The value of \(k_s\) is then obtained by enforcing the logarithmic relationship in Eq.~\ref{eq1}. High-fidelity numerical simulations, such as direct numerical simulation (DNS) or wall-resolved large-eddy simulation (LES), can also provide \(k_s\) estimates by explicitly resolving the surface roughness geometry and sampling the wall shear stress \citep{Lee2011,Cardillo2013,Olmeda2022}. However, both numerical and experimental approaches are computationally/resource expensive and often impractical, particularly for large-scale or complex systems such as atmospheric boundary layers over heterogeneous terrain \citep{Weber1999} or the roughness of biofouled ship hulls \citep{Oliveira2018}.
A significant issue with current rough flow studies, particularly for larger models, is the methodology used to scale laboratory data to real-life applications. Currently, experiments are carried out to measure $C_f$, from which $k_s$ is obtained using methods that assume homogeneous fully rough conditions such as the one in \citet{Monty2016}. These $k_s$ values are then used as input to RANS simulations to scale up the problem and obtain $C_f$ for real-life models. This approach leads to extremely high errors due to wrongfully applied assumptions in determining $k_s$ that are further amplified by current wall models in RANS \citep{Schultz2007,Schultz2021}. 

To address this challenge, we propose a new method for studying TBLs over homogeneous rough walls. This approach uses low-fidelity RANS of a TBL over a smooth wall combined with sparse PIV velocity field data of a TBL over a rough wall. It extracts secondary variables, such as \( u_\tau \), and determines roughness effects represented by \( k_s \) as a measure of the difference between the RANS set-up and PIV data. This is achieved in a cost-effective and straightforward manner, providing an accessible framework for studying TBLs over rough surfaces and advancing our understanding of roughness effects.  

The objectives of this study can be split into three parts. First, we aim to refine current wall function models to improve the reconstruction of near-wall quantities in RANS and also use them as physical constraints in the assimilation. Second, we seek to demonstrate that the data assimilation framework can effectively adjust a smooth-wall RANS baseline to match the mean flow properties of a rough-wall TBL using experimental data, while also enabling the estimation of secondary parameters even when only sparse measurements are available. Lastly, we show the current capabilities of the new wall function in RANS by using the $k_s$ from the assimilation step and monotonically increasing the Reynolds number to compare the predicted $C_f$ behaviour with the experimental data available. This approach provides a practical, cost-effective tool for characterising rough-wall flows without the need for extensive experimental or numerical campaigns.  

The rest of the paper is organised as follows. We describe the assimilation framework and set-up in \S 2. Specifically, we explain the variational method in \S 2.1, the new RANS wall function and the design variable placement for wall-bounded assimilation in \S 2.2, the computational setup in \S 2.3, a list of the cases used for the assimilation in \S 2.4, and some sensitivity studies to test the robustness of the framework in \S 2.5. Next, we go through the main results in \S 3. We start in \S 3.1 by discussing the results of the sensitivity studies. In, \S 3.2, we show the results of assimilating different roughness types, and finish in \S3.3 with analysis of the predictive capabilities of the new RANS wall-function. Finally, we present conclusions and propose directions for future work in \S 4.

\section{Methods}\label{sec2}
This section will discuss the data assimilation framework in detail, the new RANS rough-wall wall function, the wall function placement of the design variable, the OpenFOAM/DAFoam setup, a full list of cases used for this study and a brief introduction to the sensitivity tests performed to test the robustness of the framework. 

\subsection{Data assimilation framework}
A variational method is employed to combine sparse experimental data with a RANS framework. This framework minimises a cost function \(\mathcal{J}\) that measures the difference between the numerical and experimental mean velocity fields
\begin{equation}
\mathcal{J} = \frac{1}{2} \frac{\| \mathbf{U}_\text{exp} - \mathbf{U}(\mathbf{w},\mathbf{f}) \|^{2}}{\mathcal{J}_0},
\label{eq5}
\end{equation}
where \(\mathbf{U}_\text{exp}\) is the experimental mean velocity field and \(\mathbf{U}(\mathbf{w},\mathbf{f})\) is the RANS mean velocity field computed using the RANS equations. The variable \(\mathbf{w}\) represents the state variables, i.e. velocity and pressure, and \(\mathbf{f}\) the design variables, i.e. those that are tuned to minimise \(\mathcal{J}\) which is normalised by its initial value \(\mathcal{J}_0\).

Variational data assimilation is formulated as an optimisation problem, where the design variable is updated to minimise the cost function subject to constraints, given by 
\begin{align}
\min \mathcal{J}(\mathbf{w},\mathbf{f}) &, \label{eq6} \\
\text{s.t.} \; R(\mathbf{w},\mathbf{f}) =& 0, \label{eq6b}\\
\mathbf{f}_L \leq \mathbf{f} \leq \mathbf{f}_U &, \label{eq6c}
\end{align}
where \(\mathbf{f}_L\) and \(\mathbf{f}_U\) denote the lower and upper bounds on the design variable \(\mathbf{f}\), and $R(\mathbf{w},\mathbf{f})$ represents the RANS flow equations that must be satisfied.

The non-linear, constrained optimisation problem in Eq.~\ref{eq6} is solved using the discrete adjoint method. To do so, we use the open-source package DAFoam (Discrete-Adjoint for OpenFOAM) \citep{He2020}, which combines the primal solvers for the RANS equations from OpenFOAM with open-source optimisers through a python interface. The gradient of the cost function $\mathcal{J}$ with respect to the design variable, which is referred to as the sensitivity, is computed using
\begin{equation}
\frac{d\mathcal{J}}{d\mathbf{f}} = \frac{\partial \mathcal{J}}{\partial \mathbf{f}} + \psi^{T} \frac{\partial R}{\partial \mathbf{f}},
\label{eq7}
\end{equation}
where \(\psi\) is the adjoint vector.

In data assimilation, the design variable is the parameter actively tuned to minimise the discrepancy between experimental observations and numerical fields, as formulated in Eq. \ref{eq5}-\ref{eq7}. This variable can be implemented within the RANS governing equations in various ways, depending on the specific features of the flow being assimilated. A common approach is to introduce it as an additional momentum source term $\mathbf{f}_v$ \citep{Foures2014,Symon2017,Franceschini2020}, i.e.
\begin{equation}\mathbf{U} \cdot \mathbf{\nabla} \mathbf{U} + \frac{1}{\rho} \mathbf{\nabla} \overline{p} - \mathbf{\nabla} \cdot \left[ \nu_t  (\mathbf{\nabla} \mathbf{U} + \mathbf{\nabla} \mathbf{U}^T) \right] = \mathbf{f}_v,
\label{eq8}
\end{equation}
where $\overline{p}$ is the mean pressure, $\rho$ is the fluid density and $\nu_t$ is the total viscosity. This approach allows the framework to adjust the local momentum balance to reconcile differences between the RANS solution and experimental observations. However, a significant weakness of this placement is that $\mathbf{f}_v$ is not highly constrained by the underlying physics \citep{Cato2023}. Consequently, the optimiser may adjust the source term arbitrarily to match the velocity profiles, often at the expense of the eddy viscosity field. In such cases, $\nu_t$ suffers because it is partially ``absorbed" into the $\mathbf{f}_v$ term alongside corrections for 2D divergence and pressure gradient errors \citep{Padmanaban2025}. This creates a problem for the recovery of secondary variables that rely on an accurate and physically consistent $\nu_t$ field.

\subsection{Rough-wall wall functions and choice of design variable}
In a RANS framework, the effects of surface roughness are typically incorporated through modifications of the wall function, which adjusts the near-wall eddy viscosity. When the first cell centroid of the grid is not located in the viscous sublayer but instead lies within the logarithmic region, the wall shear stress, $\tau_w$, cannot be directly resolved from the near-wall velocity gradient,$\frac{\partial \mathrm{U}}{\partial \mathrm{y}}|_{@1}$ (where the subscript identifies the gradient at the first cell centroid). In this case, a wall eddy viscosity term, \(\nu_{t,w}\), is introduced to correctly enforce the wall shear stress relation
\begin{equation}
\frac{\tau_w}{\rho} = u_\tau^2 = \nu_{t,w}\frac{\partial \mathrm{U}}{\partial \mathrm{y}}|_{@1} = \nu_{t,w} \frac{\mathrm{U}_1}{\mathrm{y}_1},
\label{eq2}
\end{equation}
where \(\nu_{t,w}\) is chosen to ensure consistency with the logarithmic velocity profile.

For a smooth-wall configuration where the first cell centroid lies in the log layer, \(\nu_{t,w}\) is expressed as
\begin{equation}
    \nu_{t,w} = \frac{\nu\; \mathrm{y}_1^+}{\frac{1}{\kappa}\ln(\mathrm{y}_1^+)+B_S} = \frac{\nu\; \mathrm{y}_1^+}{\frac{1}{\kappa}\ln(E\; \mathrm{y}_1^+)},
    \label{eq4}
\end{equation}
where \(y_1^+\) denotes the non-dimensional wall-normal coordinate of the first cell centroid, defined as \(\mathrm{y}_1^+ = u_\tau \mathrm{y}_1 / \nu\), and $E = \exp(\kappa\; B_s)$. For rough walls, the formulation becomes

\begin{equation}
    \nu_{t,w} = \frac{\nu\;\mathrm{y}_1^+}{\frac{1}{\kappa}\ln(\mathrm{y}_1^+) + B_s -\Delta U^+} = \frac{\nu \; \mathrm{y}_1^+}{\frac{1}{\kappa}\ln(E'\; \mathrm{y}_1^+)},
    \label{eq3}
\end{equation}
where $E' = E/\exp(\kappa\;\Delta U^+)$.
    
Traditional RANS models for rough walls typically employ an empirical velocity scale, $u^*$, in their $\nu_t$ wall functions. This variable is analogous to $u_\tau$ but is computed empirically through the relation $u^* = C_\mu^{1/4} \sqrt{k}$, where $C_\mu$ is a constant and $k$ is the turbulent kinetic energy. This formulation presents complications since $k$ is not always explicitly computed by all eddy viscosity models, e.g., the Spalart–Allmaras (SA) model, and the reliance on $u^*$ often results in an incorrect prediction of skin friction. In contrast, advanced smooth-wall functions, such as the Spalding wall function, utilise iterative procedures to compute $u_\tau$ at any distance from the wall by employing a blending profile for U$^+$ as a function of y$^+$. To address these limitations, we propose a new wall function for rough-wall TBL RANS simulations. We adopt the iterative concept from the Spalding wall function to compute $u_\tau$ from $\nu_t$ and vice versa, but remove the blending function since, in the presence of roughness, the first cell centroid is typically already located within the logarithmic region.

By comparing Eq. \ref{eq4} and \ref{eq3}, we can see that the difference between smooth- and rough-wall behaviour can be captured by analysing $\nu_{t,w}$. This value can then be used to find $k_s$, using the logarithmic relation in Eq. \ref{eq1}. Consequently, we propose to embed the design variable in our data assimilation, hereafter referred to as $\eta$, directly within the wall function. The smooth-wall wall function is rewritten as
\begin{equation}
\nu_{t,w} = \frac{\nu \;\mathrm{y}_1^+}{\frac{1}{\kappa}\ln(E \;\mathrm{y}_1^+ / \eta)},
\label{eq10}
\end{equation}
and the rough-wall formulation as
\begin{equation}
\nu_{t,w} = \frac{\nu \;\mathrm{y}_1^+}{\frac{1}{\kappa}\ln(E' \;\mathrm{y}_1^+ / \eta)}.
\label{eq11}
\end{equation}
By moving the forcing from the outer flow to the wall boundary, we constrain the assimilation to represent roughness effects through the boundary condition itself.

While the iterative procedure currently requires the first cell centroid to be located within the logarithmic region to satisfy the log-law assumptions, this requirement is easily met in most practical applications without necessitating the high resolution required by wall-resolved methods. Furthermore, the variational method utilised here is not heavily affected by mesh size or experimental data density, meaning the framework can easily operate on coarse grids as long as the primal solver converges \citep{Thompson2024}. Therefore, having the first cell centroid in the log region is a strategic advantage that simplifies the numerical setup while ensuring that the assimilated $\nu_t$ field maintains its physical relevance for the robust estimation of $u_\tau$ and $k_s$. The complete code for this iterative wall function and the assimilation framework is provided as supplementary material.

\subsection{OpenFOAM/DAFoam set-up}
\begin{figure}[ht]
    \centering
    \includegraphics{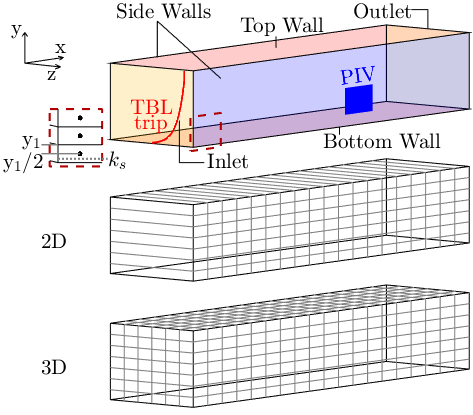}
    \caption{DAFoam set-up with bottom smooth wall, top zero-gradient wall, tripped 2cm high TBL at inlet, pressure outlet and front and back symmetry patches. The 2D domain only has 1 cell in the z-direction, while the 3D domain has $n$-cells and solves for the z-component as well}
    \label{fig3}
\end{figure}

In all cases, a consistent domain structure was utilised for the assimilation to match the experimental conditions. A simplified schematic of the domain can be seen in Fig.~\ref{fig3}. The streamwise, wall-normal and spanwise directions are denoted by x, y and z and the corresponding mean velocities are U, V and W. The domain length was set to correspond to the specific experimental facility used in the respective studies. The boundary conditions were defined as follows: the bottom boundary was modelled as a smooth wall utilising the new wall function in Eq.~\ref{eq10}; the top boundary was set to a zero-gradient condition to allow through-flow and prevent the establishment of an artificial pressure gradient; a pressure-outlet was applied at the exit; and the front and back patches were defined as symmetry planes. To reflect realistic experimental conditions, the TBL was tripped at the inlet to a height of 2 cm, ensuring the development of a representative turbulent profile.

The primal flow was solved using the steady RANS equations, with turbulence closure provided by the SA model. The SA model was selected for its robustness in boundary layer flows and its computational efficiency within the adjoint framework \citep{Spalart1994}. Numerical discretisation was performed using a second-order accurate finite volume scheme. Specifically, a second-order upwind scheme was utilised for the convection terms to minimise numerical diffusion, while the pressure-velocity coupling was handled via the SIMPLE (Semi-Implicit Method for Pressure Linked Equations) algorithm. The simulations were considered converged when the residuals for momentum and the turbulence quantities dropped below $10^{-6}$. The use of the discrete adjoint method implemented in DAFoam ensures accurate sensitivities, which are crucial for the optimiser to work efficiently in converging the objective function \citep{Kenway2019}.

The experimental PIV data were integrated into the framework as a discrete field, constrained to the specific coordinates where the measurements were recorded. This ensures the cost function only accounts for discrepancies at the specific measurement locations. While the primary assimilations were conducted in 2D to maintain computational efficiency, a 3D assimilation was also performed to investigate whether 2D divergence errors observed in previous studies \citep{Suzuki2009, Padmanaban2025} were prominent in this type of flow. For the 3D configuration, the domain in the $z$-direction was partitioned into multiple cells, rather than the single-cell depth utilised in the 2D setup, as illustrated in Fig.~\ref{fig3}. This allowed for a direct assessment of the influence of spanwise flow variations on the recovered friction and roughness parameters.

For all assimilation cases, the permissible range of the design variable $\eta$ was constrained by predefined lower and upper bounds. These bounds were strategically defined to be broader than the expected $\Delta U^+$. Numerically, it is imperative that $\eta$ remains non-zero as it resides in the denominator of the logarithmic term in the wall function. Consequently, $\eta$ was constrained between a small positive value and a large upper limit (i.e., $1 \leq \eta \leq 1000$), providing a wide range for the optimiser while maintaining numerical stability. These bounds ensure that the framework can explore the full transition from aerodynamically smooth to fully rough regimes without encountering singular values.

To evaluate the predictive capability of the modified wall function, the optimised parameters were utilised to perform RANS simulations at higher Reynolds numbers ($Re_x$) than those present in the experimental datasets. This scaling study assesses whether the modified wall function can accurately predict the development of the TBL across a wide range of $Re_x$, a critical requirement for full-scale engineering applications such as ship hull performance.

\subsection{List of cases}
To ensure visual clarity throughout the results and discussion, a consistent symbol and colour convention is adopted. Experimental data are identified by an inverted triangle ($\nabla$) in a blue gradient, while all assimilated results are identified by a circle ($\circ$) in a copper gradient. Simulation results are represented by lines in a red gradient. These cases are summarised in Table~\ref{tab1} in the order that they appear in the subsequent analysis, from darkest to lightest shade.

\definecolor{blue1}{rgb}{0.05, 0.10, 0.25}
\definecolor{blue2}{rgb}{0.12, 0.27, 0.50}
\definecolor{blue3}{rgb}{0.20, 0.45, 0.75}
\definecolor{blue4}{rgb}{0.45, 0.67, 0.87}
\definecolor{blue5}{rgb}{0.70, 0.90, 1.00}

\definecolor{cop1}{rgb}{0.2402, 0.1501, 0.0956} 
\definecolor{cop2}{rgb}{0.4853, 0.3033, 0.1931}
\definecolor{cop3}{rgb}{0.7304, 0.4565, 0.2907}
\definecolor{cop4}{rgb}{0.9755, 0.6096, 0.3882}
\definecolor{cop5}{rgb}{1.0000, 0.7628, 0.4858}

\definecolor{gre1}{rgb}{0.1922, 0.1922, 0.1922} 
\definecolor{gre2}{rgb}{0.3882, 0.3882, 0.3882}
\definecolor{gre3}{rgb}{0.5843, 0.5843, 0.5843}
\definecolor{gre4}{rgb}{0.7804, 0.7804, 0.7804}

\definecolor{red1}{rgb}{0.4167, 0     , 0} 
\definecolor{red2}{rgb}{0.9375, 0     , 0}
\definecolor{red3}{rgb}{1.0000, 0.4583, 0}

\begin{table}[ht]
\centering
\caption{Summary of experimental, assimilated and simulated cases with corresponding symbols and colour coding}
\label{tab1}
\begin{tabular}{lllc}
\toprule
Category & Study & Freestream Velocity / Case / Details & Symbol \\
\midrule
Experimental ($\nabla$) & Gul et al. (2021) & 8 ms$^{-1}$ P24 & \tikz\draw[blue1,fill=blue1] (0.5ex,0.5ex) -- (0ex,-0.5ex) -- (-0.5ex,0.5ex) -- cycle; \\
& Gul et al. (2021) & 18 ms$^{-1}$ P24 & \tikz\draw[blue2,fill=blue2] (0.5ex,0.5ex) -- (0ex,-0.5ex) -- (-0.5ex,0.5ex) -- cycle; \\
& Gul et al. (2021) & 22 ms$^{-1}$ P24 & \tikz\draw[blue3,fill=blue3] (0.5ex,0.5ex) -- (0ex,-0.5ex) -- (-0.5ex,0.5ex) -- cycle; \\
& Medjnoun et al. (2023) & 10 ms$^{-1}$ Config 1 & \tikz\draw[blue4,fill=blue4] (0.5ex,0.5ex) -- (0ex,-0.5ex) -- (-0.5ex,0.5ex) -- cycle; \\
& Wangsawijaya et al. (2023) & 20 ms$^{-1}$ R1 & \tikz\draw[blue5,fill=blue5] (0.5ex,0.5ex) -- (0ex,-0.5ex) -- (-0.5ex,0.5ex) -- cycle; \\
\midrule
Assimilated ($\circ$) & Gul et al. (2021) & 8 ms$^{-1}$ P24 & \tikz\draw[cop1,fill=cop1] (0,0) circle (0.5ex); \\
& Gul et al. (2021) & 18ms$^{-1}$ P24 & \tikz\draw[cop2,fill=cop2] (0,0) circle (0.5ex); \\
& Gul et al. (2021) & 22 ms$^{-1}$ P24 & \tikz\draw[cop3,fill=cop3] (0,0) circle (0.5ex); \\
& Medjnoun et al. (2023) & 10 ms$^{-1}$ Config 1  & \tikz\draw[cop4,fill=cop4] (0,0) circle (0.5ex); \\
& Wangsawijaya et al. (2023) & 20 ms$^{-1}$ R1 & \tikz\draw[cop5,fill=cop5] (0,0) circle (0.5ex); \\
\midrule
Simulated ($-$) & Gul et al. (2021) &  P24, $Re_x$-sweep& \tikz\draw[red1,fill=red1] (0.75ex,0.25ex) -- (0.75ex,-0.25ex) -- (-0.75ex,-0.25ex) -- (-0.75ex,0.25ex) -- cycle; \\
& Medjnoun et al. (2023) & Config 1, $Re_x$-sweep & \tikz\draw[red2,fill=red2] (0.75ex,0.25ex) -- (0.75ex,-0.25ex) -- (-0.75ex,-0.25ex) -- (-0.75ex,0.25ex) -- cycle; \\ 
& Wangsawijaya et al. (2023) & R1, $Re_x$-sweep & \tikz\draw[red3,fill=red3] (0.75ex,0.25ex) -- (0.75ex,-0.25ex) -- (-0.75ex,-0.25ex) -- (-0.75ex,0.25ex) -- cycle; \\
\bottomrule
\end{tabular}
\end{table}

\subsection{Sensitivity tests}
In addition to the baseline cases, a series of sensitivity tests was performed to evaluate the robustness of the assimilation framework. To begin with, a baseline simulation was initialised with incorrect roughness properties. The purpose of this test, referred to as the ``wrong $k_s$'' baseline, is to verify that the assimilation can correct the RANS baseline when its initial condition is an incorrect rough-wall state rather than the standard smooth-wall baseline. In other words, the assimilation should not be sensitive to initial conditions. Secondly, the sensitivity of the framework to experimental data density was investigated. While the primary results utilise a wider window of PIV data, an additional assimilation was conducted using only a single velocity profile sampled in the middle of the window of available PIV data. These tests aim to establish the minimum data requirements necessary for the framework to accurately recover $u_\tau$ and $k_s$ and to assess the resilience of the variational method when provided with sparse experimental input. Lastly, the assimilation was tested on a 3D domain to investigate whether this type of flow has any non-negligible 3D effects that result in high 2D divergence errors.

In the “wrong $k_s$” baseline check, the bounds were set sufficiently low to ensure the framework could capture the correct $\Delta U^+$ from an arbitrary “rougher” initial guess (i.e. $0.01 \leq \eta \leq 100$). For the other tests, we used the same bounds as the cases listed in Table~\ref{tab1}, i.e. $1 \leq \eta \leq 1000$. Similarly to \S2.4, a standard symbol and colour convention was used for the sensitivity tests, as shown in Table~\ref{tab2}.

\begin{table}[ht]
\centering
\caption{Summary of sensitivity tests conducted with the \cite{Gul2021} P24 18ms$^{-1}$ case with corresponding symbols and colour coding}
\label{tab2}
\begin{tabular}{llc}
\toprule
Category & Details & Symbol \\
\midrule
Experimental ($\nabla$) & Gul et al. (2021) - P24 18ms$^{-1}$ & \tikz\draw[blue2,fill=blue3] (0.5ex,0.5ex) -- (0ex,-0.5ex) -- (-0.5ex,0.5ex) -- cycle; \\
\midrule
Assimilated ($\circ$)& 2D assimilation, smooth wall baseline, downsampled exp. dataset & \tikz\draw[gre1,fill=gre1] (0,0) circle (0.5ex); \\
& 2D assimilation, smooth wall baseline, single exp. profile & \tikz\draw[gre2,fill=gre2] (0,0) circle (0.5ex); \\
& 2D assimilation, wrong $k_s$ baseline,  downsampled exp. dataset & \tikz\draw[gre3,fill=gre3] (0,0) circle (0.5ex); \\
& 3D assimilation, smooth wall baseline, downsampled exp. dataset & \tikz\draw[gre4,fill=gre4] (0,0) circle (0.5ex); \\
\bottomrule
\end{tabular}
\end{table}

\section{Results}\label{sec3}
The results are presented by first examining the outcomes of the sensitivity tests to establish the framework's robustness, followed by an analysis of the primary results across the various roughness datasets.

\subsection{Sensitivity tests results}
The performance of the data assimilation framework is first evaluated through the convergence of the objective function, $\mathcal{J}$, for the sensitivity cases listed in Table~\ref{tab2}. As the placement of the design variable within the wall function is inherently sensitive to overfitting, a rigorous sampling strategy was implemented to ensure the physical validity of the recovered parameters. Specifically, the optimal state for each case was sampled only when the following three criteria were simultaneously satisfied:

\begin{enumerate}
\item The objective function reached a stable plateau, indicating a minimised discrepancy between the RANS solution and the PIV data.
\item The assimilated streamwise velocity, $\mathbf{U}$, at iteration $t$ was within 5\% of the experimental values.
\item The change in assimilated velocity in the region where experimental data was provided between successive iterations, $\Delta \mathbf{U} = |\mathbf{U}_{t+1} - \mathbf{U}_{t}|$, was less than 1\%.
\end{enumerate}
These criteria prevent numerical artefacts and ensure that the final $\eta$ value represents a stable momentum balance. 

\begin{figure}[ht]
\centering
    \includegraphics{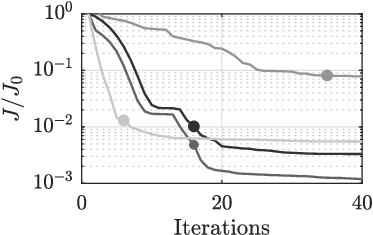}
    \caption{Objective function convergence for sensitivity tests, the colour map used is explained in Table \ref{tab2}}
    \label{fig4}
\end{figure}

\begin{figure}[ht]
\begin{subfigure}[b]{.55\textwidth}
    \includegraphics{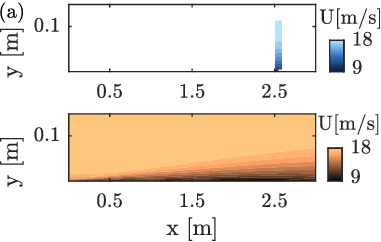}
\end{subfigure}\begin{subfigure}[b]{.45\textwidth}
    \includegraphics{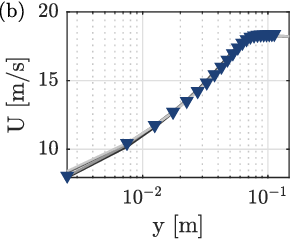}
\end{subfigure}
\caption{(a) Mean streamwise velocity contour plot with top panel: experimental data used in the assimilation from \citet{Gul2021} - case P24 18ms$^{-1}$ - and bottom panel: assimilated field. (b) Streamwise averaged mean velocity profiles from all sensitivity tests listed in Table~\ref{tab2}}
\label{fig5}
\end{figure}

The convergence history for the sensitivity tests is illustrated in Fig.~\ref{fig4}. It is important to note that the experimental data used for these sensitivity tests, along with all other datasets from \citet{Gul2021}, were affected by heavy laser reflections in the bottom-right corner of the PIV data. To maintain data integrity, a significant portion of the experimental data were removed in this region, as shown in the top panel of Fig.~\ref{fig5}a. The sampling point for each case is identified by a circle ($\circ$) superimposed on the convergence lines, with the colour coding maintained as per the convention established in Table~\ref{tab2}. Despite the variations in initial conditions and experimental datasets, all cases demonstrate a monotonic decrease in the objective function, with the 3D case converging to a higher value and with a more shallow descent in comparison to the other cases. These observations are consistent with the 3D assimilations performed in \citet{Padmanaban2025}.

The assimilated field, shown in the bottom panel of Fig.~\ref{fig5}a, demonstrates how the framework ``fixes'' the PIV field by physically reconstructing the flow in the missing region. This reconstruction is further validated in Fig.~\ref{fig5}b, which compares the streamwise-averaged velocity profiles (in physical units) from all sensitivity tests against the experimental profiles from \citet{Gul2021}. The profiles are averaged across the window of available experimental data, and the close agreement confirms that the framework maintains physical consistency even when a portion of the measurement field is discarded.

\begin{figure}[ht]
\begin{subfigure}[b]{.5\textwidth}
    \includegraphics{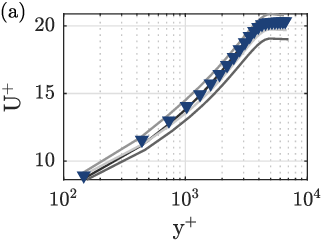}
\end{subfigure}\begin{subfigure}[b]{.5\textwidth}
    \includegraphics{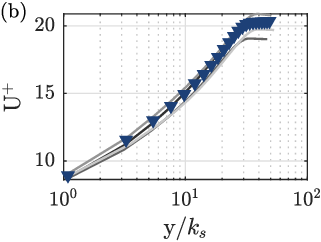}
\end{subfigure}
\caption{Sensitivity tests results from Table~\ref{tab2}. Comparison between experimental and assimilated streamwise-averaged mean velocity in viscous units plotted against the wall-normal coordinate in viscous units (a) and against the wall-normal coordinate normalised by $k_s$ (b)}
\label{fig6}
\end{figure}

As previously mentioned in \S 2.2, the reconstruction of the velocity field is driven by the embedding of the design variable directly within the wall function. By imposing these physical constraints, the assimilation framework is forced to determine the correct eddy viscosity distribution required to satisfy the momentum balance. It is this recovered eddy viscosity that allows for the accurate subsequent recovery of the friction velocity, $u_\tau$, as defined by Eq.~\ref{eq10} and \ref{eq11}. As shown in Fig.~\ref{fig6}a, nearly all sensitivity cases demonstrate an excellent recovery of $u_\tau$. Initiating the assimilation from a baseline with an incorrect $k_s$ yielded results identical to the smooth-wall baseline, with both recovering $u_\tau$ to within 2\% of the experimental value. The 3D assimilation yielded results nearly identical to the 2D configuration ($u_\tau<3\%$ off the experimental value), confirming that 3D constraints are unnecessary for this specific flow regime and justifying the use of more computationally efficient 2D models. The only notable discrepancy occurred in the case utilising a single velocity profile, which resulted in a 6\% difference. Nonetheless, it reinforces the ability of assimilation to reconstruct the TBL even with extremely sparse input data.

Finally, the framework's performance in determining the equivalent sand-grain roughness, $k_s$, is evaluated in Fig.~\ref{fig6}b. The profiles, normalised by the recovered roughness height (y$/k_s$), show a high degree of fidelity across the test suite, confirming that the design variable effectively captured the roughness effects within the wall function. The highest relative discrepancy for $k_s$ was observed in the 3D assimilation at approximately 8\%; however, this corresponds to an absolute difference of less than 0.2 mm, which is therefore negligible in a physical context.

In summary, the sensitivity tests demonstrate that the assimilation framework is highly robust to varying initial conditions, as the choice of a smooth or rough baseline does not influence the final recovered state. The framework proves capable of ``fixing'' problematic regions in PIV datasets, such as those obscured by reflections or uneven lighting, and remains functional even with extremely sparse data, such as a single velocity profile. Finally, the results indicate that 3D constraints are not necessary for this specific flow regime.

\subsection{Assimilation of different roughness types}

Following the sensitivity tests, the framework was applied to the remaining datasets using the smooth-wall baseline 2D setup. The full list of experimental cases investigated in this study is introduced in \S 2.3, and the colour scheme established in Table~\ref{tab1} is maintained for all subsequent plots to ensure consistency.

\begin{figure}[ht]
\begin{subfigure}[b]{.55\textwidth}
    \includegraphics{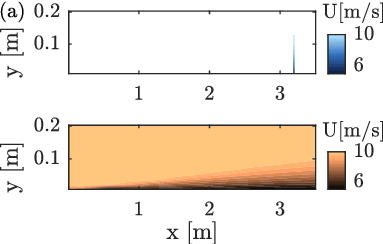}
\end{subfigure}\begin{subfigure}[b]{.45\textwidth}
    \includegraphics{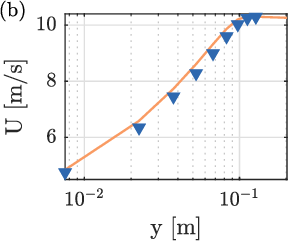}
\end{subfigure}
\begin{subfigure}[b]{.55\textwidth}
    \includegraphics{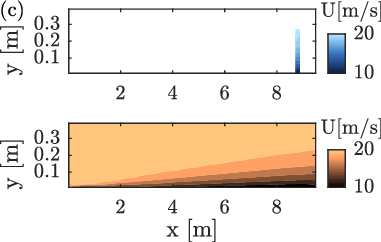}
\end{subfigure}\begin{subfigure}[b]{.45\textwidth}
    \includegraphics{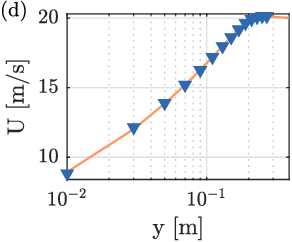}
\end{subfigure}
\caption{Mean streamwise velocity contour plots. Top panel: experimental data used in the assimilation from (a) \citet{Medjnoun2023} case 10ms$^{-1}$ and (c) \citet{Wangsawijaya2023} case R1 20ms$^{-1}$. Bottom panel: assimilated field. Streamwise-averaged mean velocity profiles from (b) \citet{Medjnoun2023} and (d) \citet{Wangsawijaya2023}}
\label{fig7}
\end{figure}

The assimilated and experimental mean velocity contours for the remaining two studies are presented in Fig.~\ref{fig7}a and \ref{fig7}c. In these plots, the top panels represent the experimental data fields, while the bottom panels display the corresponding assimilated velocity fields. Contour plots for the additional cases in \citet{Gul2021} have been omitted, as they are qualitatively identical to Fig.~\ref{fig5}a. For all cases in \citet{Gul2021}, a section of the bottom-right corner of the experimental data was removed as discussed in \S 3.1. Additionally, the \citet{Medjnoun2023} case in Fig.~\ref{fig7}a was captured within a limited spatial window. After downsampling to match the CFD grid, this resulted in only two available velocity profiles and was further restricted to the streamwise component alone, whereas the other datasets contained two velocity components. In contrast, the data from \citet{Wangsawijaya2023} in Fig.~\ref{fig7}c represents the most complete dataset provided to the framework. 

Despite these differences in data availability and quality, the contour plots exhibit consistent behaviour across all assimilations. The TBL growth from the inlet to the measurement location is clearly captured, with no unphysical behaviour observed upstream, downstream, or above the measurement window. This confirms that the physical constraints on the design variable effectively isolated the differences between the smooth-wall baseline and the rough-wall experimental data. In turn, the remainder of the flow field was solely determined by the RANS governing equations and the turbulence model, ensuring that the results remain physically consistent across the entire domain. This is further supported by Fig.~\ref{fig7}b and \ref{fig7}d, where the streamwise-averaged mean velocity profiles across the region of available data from the experiment and the assimilation are compared. These plots provide a quantitative demonstration of how precisely the assimilation matches the experimental data.

\begin{figure}[ht]
\begin{subfigure}[b]{.5\textwidth}
    \includegraphics{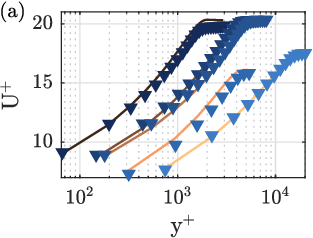}
\end{subfigure}\begin{subfigure}[b]{.5\textwidth}
    \includegraphics{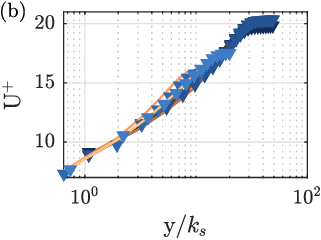}
\end{subfigure}
\caption{Comparison between experimental and assimilated mean streamwise velocity in viscous units plotted against the wall normal coordinate (a) in viscous units and (b) normalised by the equivalent sand-grain roughness height. The colour maps are defined in Table \ref{tab1}}
\label{fig8}
\end{figure}

Moving away from the qualitative contour analysis and wall-normal profiles in physical units, Fig.~\ref{fig8}a shows the mean streamwise velocity profiles, averaged in the streamwise direction within the window of available experimental data, plotted in viscous units (U$^+$ vs y$^+$). These profiles provide a direct measure of how effectively the friction velocity, $u_\tau$, was recovered by the assimilation framework. As shown in Fig.~\ref{fig8}a, all cases demonstrate an excellent recovery of $u_\tau$, with the reconstructed values falling within 3\% of the experimental measurements. An interesting result is observed in the \citet{Medjnoun2023} case. Despite the experimental data being limited to only two profiles, the reconstruction yielded results comparable to those where the entire downsampled field was utilised—a marked improvement over the single-profile sensitivity test.

Having established that the assimilation framework successfully recovers $u_\tau$, it is necessary to evaluate its performance in determining the equivalent sand-grain roughness, $k_s$, which is a primary parameter in roughness studies. Fig \ref{fig8}b shows the mean velocity profiles for all cases, again averaged in the streamwise direction within the window of available experimental data, plotted in viscous units (U$^+$) against the wall-normal coordinate normalised by the recovered roughness (y$/k_s$). In roughness research, this scaling is used to identify self-similar profiles within the logarithmic region. Collapsed profiles indicate that the TBL is in equilibrium and the roughness is effectively homogeneous. For the purpose of this study, this plot confirms the accuracy of the $k_s$ recovery. As shown in Fig.~\ref{fig8}b, the recovered $k_s$ values for all baseline cases are within 3\% of the published experimental values. The most accurate reconstruction was achieved for the \citet{Medjnoun2023} case, which fell within less than 1\% of the reported value.

\subsection{Predictive capabilities of the RANS wall function}
\begin{figure}[ht]
\centering
\includegraphics{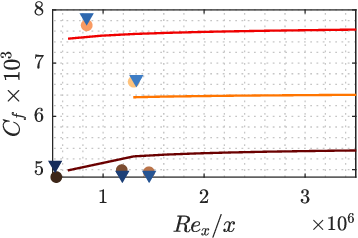}
\caption{Friction coefficient plotted against Reynolds number, $Re_x$. Comparison between experimental, assimilated and simulated data. The colour maps are defined in Table~\ref{tab1}}
\label{fig9}
\end{figure}

The final stage of this study evaluates the predictive capability of the modified wall function. To assess this, the $k_s$ values obtained from the assimilation were utilised as fixed inputs for the wall function, rather than relying on the experimental data directly. A series of simulations was conducted across a range of Reynolds numbers to generate $C_f$ vs. $Re_x$ curves, scaling up from the initial assimilated conditions. This approach addresses a known shortcoming of standard RANS models in predicting rough-wall scaling, as discussed in \S 1. The results of these simulations are presented in Fig.~\ref{fig9}, where the assimilated and experimental markers follow the conventions established in Table~\ref{tab1}.

As shown in Fig.~\ref{fig9}, the predicted $C_f$ values for the \citet{Medjnoun2023} and \citet{Wangsawijaya2023} cases fall within $\approx5\%$ of the experimental measurements. Furthermore, the friction remains constant at higher $Re_x$, which is consistent with the expected behaviour of a fully rough flow sampled at a constant x with varying U$_\infty$. The most significant discrepancy is observed in the \citet{Gul2021} case ($\approx 9\%$), which represents the smallest roughness type across all datasets. It should be noted that even a minute absolute difference in $u_{\tau}$ results in a disproportionally higher relative difference in $C_f$. By analysing these results, the primary conclusion is that the accuracy of the assimilation and the subsequent RANS prediction step is fundamentally dependent on the quality and clarity of the input experimental data. While the framework demonstrates robust scaling for high-quality datasets, it also highlights that the reliability of the recovered parameters is inherently linked to the precision of the original measurements.

\section{Conclusions}\label{sec4}
This study has presented and validated a data assimilation framework designed to recover friction and roughness parameters from experimental TBL data. By embedding a design variable directly within a reformulated RANS wall function, the framework successfully bridges the gap between smooth-wall baseline simulations and rough-wall experimental data. The results demonstrate that the framework is highly effective at reconstructing both the friction velocity and the equivalent sand-grain roughness, with the majority of cases showing a recovery accuracy within 3\% of published experimental values. This low discrepancy was maintained even when the assimilation was initiated from a baseline with an incorrect roughness height, indicating that the optimiser is insensitive to the initial guess of the surface state.

The framework demonstrates robustness to varying data quality, with sensitivity analysis showing that just two velocity profiles provide accuracy comparable to a full PIV field. It effectively resolves flow in obscured regions and confirms that 2D assimilation is sufficient for this type of boundary layers, avoiding unnecessary 3D computational costs. Furthermore, predictive simulations at higher Reynolds numbers validate that recovered $k_{s}$ values accurately extrapolate results to full-scale conditions. Predictions for TBLs measured by \citet{Medjnoun2023} and \citet{Wangsawijaya2023} fell within 5\% of measured 
$C_{f}$ values, capturing the expected constant friction behaviour. Meanwhile, larger discrepancies in the \citet{Gul2021} cases indicate that assimilation results reflect the initial experimental uncertainty.

Ultimately, this framework serves as a versatile tool for enhancing experimental data, providing a robust method to obtain secondary variables, such as friction and roughness scales, and also resolving issues in datasets affected by uneven lighting or laser reflections. Moreover, when used in conjunction with RANS, it also provides the predictive capability required to obtain skin friction information for higher $Re_x$ or larger-scale models. Future work will focus on extending this framework to heterogeneous surfaces using the datasets in \citet{Formichetti2025}.

\backmatter

\bmhead{Supplementary information}
The OpenFOAM source code for the modified wall function, along with the implementation of the adjoint-based design variable for DAFoam, is available as supplementary material accompanying this manuscript.

\bmhead{Acknowledgements}
The authors acknowledge funding from the European Office for Air Force Research and Development (Grant ref: FA8655-23-1-7005) and EPSRC (Grant ref no: EP/W026090/1).

\bmhead{Conflict of interest}
The authors report no conflict of interest.

\bmhead{Data availability}
All data published in this article is available as supplementary material accompanying this manuscript.

\bmhead{Authors ORCID} M. Formichetti, https://orcid.org/0000-0003-4450-334X; U. Cadambi Padmanaban https://orcid.org/0000-0003-2011-2838; P. He, https://orcid.org/0000-0001-5074-7215; S. Symon, https://orcid.org/0000-0001-9085-0778; B. Ganapathisubramani, https://orcid.org/0000-0001-9817-0486.

\bibliography{sn-bibliography}

\end{document}